\documentclass[10pt,aps,pra,twocolumn,noshowpacs,superscriptaddress]{revtex4}

\usepackage{mathrsfs, bm,color,amsmath,amssymb, stmaryrd, amsfonts,latexsym,graphicx, psfrag}
\usepackage{bbold}	

\newcommand{\im}{\mathrm{Im}\,}

\newcommand{\be}{\begin{equation}}
\newcommand{\ee}{\end{equation}}
\newcommand{\ba}{\begin{eqnarray}}
\newcommand{\ea}{\end{eqnarray}}

\newcommand{\pol}{\mathcal{P}}
\newcommand{\el}{\mathcal{E}}
\newcommand{\en}{\mathrm{E}}
\newcommand{\free}{\mathrm{F}}
\newcommand{\ham}{\mathrm{H}}
\newcommand{\bco}{\mathcal{A}}
\newcommand{\gpot}{\Omega}
\newcommand{\diff}{\mathrm{d}}
\newcommand{\id}{\mathbb{1}}

\newcommand{\abs}[1]{\left\vert#1\right\vert}

\newcommand{\psop}[3]{\left<#1\middle\vert#2\middle\vert#3\right>}
\newcommand{\ps}[2]{\left<#1\middle\vert#2\right>}
\newcommand{\ket}[1]{\left\vert#1\right>}
\newcommand{\bra}[1]{\left<#1\right\vert}

\newcommand{\remove}[1]{}

\renewcommand{\l}{\left}
\renewcommand{\r}{\right}

\definecolor{colorforfred}{rgb}{1,0.5,0}

\begin{document}
\title{Statistical mechanics approach to the\\ 
electric polarization and dielectric constant of band insulators}

\author{Fr\'ed\'eric \surname{Combes}}
\email{frederic.combes@u-psud.fr}
\affiliation{Laboratoire de Physique des Solides, CNRS UMR 8502, Univ. Paris-Sud, F-91405 Orsay Cedex, France}
\author{Maximilian \surname{Trescher}}
\affiliation{Dahlem Center for Complex Quantum Systems and Institut f\"ur Theoretische Physik, Freie Universit\"at Berlin, Arnimallee 14, D-14195 Berlin, Germany}
\author{Fr\' ed\' eric \surname{Pi\'echon}}
\affiliation{Laboratoire de Physique des Solides, CNRS UMR 8502, Univ. Paris-Sud, F-91405 Orsay Cedex, France}
\author{Jean-No\"el \surname{Fuchs}}
\affiliation{Laboratoire de Physique des Solides, CNRS UMR 8502, Univ. Paris-Sud, F-91405 Orsay Cedex, France}
\affiliation{Laboratoire de Physique Th\' eorique de la Mati\` ere Condens\' ee, CNRS UMR 7600, Univ. Pierre et Marie Curie 4, place Jussieu, 75252 Paris Cedex 05, France}

\date{\today}

\begin{abstract}
We develop a theory for the analytic computation of the free energy of band insulators in the presence of a uniform and constant electric field. The two key ingredients are a perturbation-like expression of the Wannier-Stark energy spectrum of electrons and a modified statistical mechanics approach involving a local chemical potential in order to deal with the unbounded spectrum and impose the physically relevant electronic filling. At first order in the field, we recover the result of King-Smith, Vanderbilt and Resta for the electric polarization in terms of a Zak phase -- albeit at finite temperature -- and, at second order, deduce a general formula for the electric susceptibility, or equivalently for the dielectric constant. Advantages of our method are the validity of the formalism both at zero and finite temperature and the easy computation of higher order derivatives of the free energy. We verify our findings on two different one-dimensional tight-binding models.
\end{abstract}


\maketitle

\section{Introduction}

From the classical point of view, a periodic distribution of discrete charges -- as found in a crystal -- possesses a well-defined polarization (i.e. an electric dipole moment) if it is charge neutral. However, due to the periodicity of the charge distribution, this polarization is only defined modulo a Bravais vector, that is, an integer in the proper units. The latter is misleadingly known as the quantum of polarization, although unrelated to Planck's constant. In addition to that, if dynamics are specified for this charge distribution, then it also possesses an electric susceptibility $\chi$, related to the dielectric constant $\epsilon$ by $\epsilon=1+\chi$. The classical picture (also known as the Clausius-Mossoti approach) \cite{Ashcroft}, however, fails when the electrons are described at the quantum level, as extended Bloch states lead to a continuous charge distribution rendering the classical formula meaningless. Hence, one  must use a new approach to describe the polarization of a crystal. Since the work of King-Smith, Vanderbilt and Resta (KVR) \cite{KSV,Resta1994,VanderbiltResta}, tools needed to compute the electric polarization in crystals are available. Their approach, named \emph{modern theory of polarization}, is based on the understanding that a change of polarization corresponds to an adiabatic flow of charges in an insulator (for a pedagogical review of the modern theory of polarization, see \cite{Spaldin}). The current resulting from an adiabatic deformation of the crystal can easily be computed from the Bloch wavefunctions, and the resulting change in polarization is proportional to the difference of the Zak phase \cite{Zak1989} between the initial and final states. In turn, the Zak phase corresponds to the position of the Wannier center of a Bloch band inside a unit cell. In light of this fact, we can use a localization prescription for the delocalized Bloch wavefunctions of a band: if we assume that the electrons (and therefore their whole charge) are localized at their Wannier centers, then the classical formula for discrete charge distribution gives the expected result for the polarization. From the Zak phase properties we gain the insight that, at the quantum level, the polarization does not depend on the density of charge (i.e. the modulus square of the Bloch states), but on the phase of the Bloch states. In this context, the quantum of polarization appears related to the Zak phase being defined modulo $2\pi$.

Despite the breakthrough of this approach, there are still difficulties with it. From a general point of view, the polarization is an equilibrium quantity -- that should be computable from standard statistical mechanics -- and not a transport property, while the modern theory of polarization relies on adiabatic currents. Also it is essentially restricted to zero temperature. Here, we propose to adopt a statistical mechanics approach similar to the one usually developed for the magnetic response (see for example \cite{Raoux2015}). Using a scalar gauge, we start by computing the energy spectrum of band electrons in an electric field (the well-known Wannier-Stark ladder \cite{Wannier1960,Zak1968,Resta2000,Niu2010}) at second order in the electric field. This spectrum is unbounded, which constitute a major difficulty for a standard statistical mechanics approach. To circumvent this difficulty, we develop a modified approach that takes into account the fact that the band insulator in a weak electric field remains translationally invariant in practice. The key ingredient is to introduce a local chemical potential that forces the physical electronic filling in each unit cell.

After the pioneering work of King-Smith and Vanderbilt \cite{KSV} and of Resta \cite{Resta1994}, there have been many further developments in the computation of the dielectric properties of insulating crystals. Here, we briefly review some of these. Nunes and Vanderbilt have developed a real-space approach to the computation of electric polarization and susceptibility \cite{NunesVanderbilt}. It is based on the KVR formula extended to treat the case of field-induced polarization and not only spontaneous polarization. Several authors have adapted density functional theory to a finite electric field by introducing an energy functional that depends on the electric polarization as computed using the KVR formula \cite{Gonze,Souza,Umari}. In this way they can access the dielectric susceptibility and higher order response functions, however the electric polarization is taken from KVR. Kirtman and co workers have developed a vector potential approach that bypasses the difficulty related to the unbounded position operator present in the scalar electric potential but has other problems \cite{Kirtman}. The same authors \cite{SpringborgKirtman} have also attacked the problem by replacing the true electric scalar potential by a piecewise linear (such as sawtooth or continuous triangular) potential that has the advantage of corresponding to a periodic Hamiltonian. A drawback of this approach is that it does not recover the KVR polarization. Recently, Nourafkan and Kotliar have included correlations effects in the computation of the electric polarization \cite{Nourafkan2013}. Swiecicki and Sipe use linear response theory at finite frequency to obtain the dielectric function \cite{SwiecickiSipe}.

In the following, we consider one-dimensional tight-binding models of band insulators as the minimal models capturing the physics at stake. The outline of the paper is as follows. In section II we give a general derivation of the electric polarization and susceptibility at finite temperature. The two crucial steps in the derivation are (i) a perturbation-like expression (power series in the electric field) for the energy spectrum of a tight-binding model in a constant electric field, and (ii) modified statistical mechanics involving a local chemical potential in order to properly handle the unbounded energy spectrum. Then, in section III, we check our results and approximations on two toy-models that can either be solved analytically or numerically. Section IV contains a discussion and a conclusion. In appendices, we give details on the derivations. Appendix A discusses the position operator. Appendix B gives the chemical potential as a function of the electric field. Appendix C shows that strong interactions between electrons impose a local electronic filling. Appendix D presents an alternative derivation of the finite temperature polarization and susceptibility based on the charge density.

\section{General derivation}

We consider a one-dimensional tight-binding model for electrons in a periodic crystal made of $N$ unit cells, each containing $N_b$ sites/orbitals. Eventually, we are interested in studying the thermodynamic limit in which $N\to \infty$ at fixed number of bands $N_b$. The ions are treated minimally; they form a rigid lattice, have no dynamics, have no atomic polarizability but they do carry an electric charge so as to make the whole system charge neutral. As the spin plays no essential role in the presence of an electric field, we assume spinless electrons that carry a charge $-e=-1$. The Fermi energy is set within a band gap, such that the whole system is a charge-neutral insulating crystal (a dielectric).

\subsection{Free energy and its field derivatives: global chemical potential}

As electric polarization and susceptibility are defined for charge neutral systems only, any statistical mechanics approach must take place in the canonical ensemble where the number of electrons is fixed. The polarization $\pol$ and susceptibility $\chi$ are defined as the first and second derivatives of the free energy $\free$ with respect to the electric field $\el$, at vanishing electric field
\be
  \pol = \l. -\frac{1}{L} \frac{\partial}{\partial \el} \free \r\vert_{\el=0}
  \label{thermopol}
\ee
and
\be
  \chi = \l. -\frac{1}{L} \frac{\partial^2}{\partial \el^2} \free \r\vert_{\el=0}
  \label{thermosusc}
\ee
where $L$ is the length of the crystal.

The total free energy $F$ contains two contributions: one due to the ions and one due to the electrons. As we wish to focus on the electronic contribution we treat the ions as static charges in a scalar potential. This choice leads to their contribution to the free energy as
\be
  \free_\textrm{ions} = \sum_{n}\sum_{i=1}^{N_b} q_i (-\el (n a+x_i))=\sum_n -q \el n a
  \label{fions}
\ee
where the sum over $n$ is a sum over the unit cells ($n$ is a unit cell index taking $N$ values), $q=\sum_{i=1}^{N_b}q_i$ is the total ionic charge in a unit cell, and the origin of position is taken as the (charge-weighted) barycenter of the ions of the $n=0$ unit cell \cite{footnote1}:
\be
\bar{x}=\frac{1}{q}\sum_{i=1}^{N_b} q_i x_i=0
\ee
In the following, we set the lattice spacing $a=1$.

For technical simplicity, we introduce a chemical potential $\mu$ and compute the free energy of electrons
\be
  \free_{e^-}\l(N_{e^-},\el,\beta\r) = \mu N_{e^-} + \gpot_{e^-}\l(\mu,\el,\beta\r)
  \label{fee}
\ee
from the grand-potential
\be
\gpot_{e^-}\l(\mu,\el,\beta\r) = - \frac{1}{\beta} \sum_{\gamma} \ln\l(1+e^{-\beta(\en_\gamma-\mu)}\r)
\label{gpot1}
\ee
where $\beta=1/T$ is the inverse temperature. The chemical potential $\mu$ is used to impose the overall (i.e. global) charge neutrality of the system and $\gamma$ is the set of quantum numbers labeling the energy spectrum. For the moment, we assume that we are able to compute the energy spectrum $\{E_\gamma\}$ of a single electron in the lattice in the presence of an electric field (this energy spectrum is discussed in the next section). The total number of electrons is
\be
  N_{e^-}=-\frac{\partial \gpot_{e^-}}{\partial \mu} =\sum_{\gamma} n_F(\en_\gamma-\mu)
  \label{ne}
\ee
where
\be
  n_F(E)=\frac{1}{e^{\beta E}+1}
\ee
is the Fermi function at zero chemical potential. Charge neutrality means that $N_{e^-}=q N$. Inverting equation (\ref{ne}) gives the chemical potential $\mu$ as a function of $\beta$, $N_{e-}$ and $\el$.

\subsection{Generalities on the Wannier-Stark ladder energy spectrum for an infinite system}

In order to conduct a statistical mechanics approach, we need to know what kind of energy spectrum we have to deal with. The case of an electric field in a tight-binding Hamiltonian is delicate as the spectrum is unbounded: there are states of infinite positive and negative energies. We consider a one-dimensional tight-binding Hamiltonian $\ham_0$ describing the electrons in a crystal, in which we introduce the electric field $\el$ using the scalar gauge $A_0=-\el X$ \cite{footnote2}. The full Hamiltonian is then
\be
  \ham = \ham_0 + \el X
\ee
where $X$ is the position operator (see Appendix \ref{positionoperator}). As $\ham_0$ is invariant under translation of one unit cell, it commutes with the translation operator $T_a$ ($a = 1$ is the size of the unit cell) such that $T_a^{-n} \ham_0 {T_a}^n = \ham_0$ for any integer $n$. 

By contrast the position operator verifies the ``ladder'' identity $T_a^{-n} X {T_a}^n = X + n\id$.  As a consequence, if $\ket{\psi}$ is an eigenvector of $\ham$ with eigenvalue $\en(\el)$, then successive application of the translation operator implies that $T_a^n\ket{\psi}$ is also an eigenvector of $\ham$ with eigenvalue $\en(\el) + n\el $. Defining $\ket{\en}$ as the eigenstate such that $-1/2\le X(\el)=\bra{\en} X \ket{\en}<1/2$ and $\en(\el)$ the corresponding eigen-energy we deduce that, for any integer $n$, the translated state 
\be
\ket{\en_n}=T_a^n\ket{\en},
\label{ketws}
\ee
is an eigen-state of energy
\be
\en_n(\el) = \en(\el) + \el n
\label{ews}
\ee
with a mean position
\be
X_n(\el)=\bra{\en_n} X \ket{\en_n}=X(\el)+n.
\label{xws}
\ee
translated by $n$ unit cells. The integer $n$ labels the unit cells (in a finite system it takes $N$ values). Eqs. (\ref{ketws}), (\ref{ews}) and (\ref{xws}) are the essential characteristics of the so-called Wannier-Stark ladder (WSL) \cite{Wannier1960}. The eigenstates are called Wannier-Stark (WS) states. A simple counting argument shows that if the tight-binding spectrum at zero field comprises $N_b$ bands, each containing $N$ states, then the full spectrum of the Hamiltonian $\ham$ is made by $N_b$ such ladders that will be labeled by an index $\alpha = 1,...,N_b$. This can be schematically written as:
\be
\left\lbrace
\begin{array}{l}
\ket{\en^{\alpha}_{n}}=T_a^n\ket{\en^{\alpha}}\\
\en^\alpha_n(\el)= \en^{\alpha}(\el) + \el n\\
X^\alpha_n(\el)=X^{\alpha}(\el)+n\\
\end{array}
\right.
\label{wsprop}
\ee
where hereafter the quantities $\ket{\en^{\alpha}}$, $\en^{\alpha}(\el)$ and $-1/2\le X^{\alpha}(\el)<1/2$ are referred to as respectively the center states, the energy center and the position center of the $\alpha$ ladder. 

The validity of properties (\ref{wsprop}) necessarily implies that the WS states $\ket{\en^{\alpha}_n}$ are localized states such that we can also associated a localization length $\xi_{\alpha}(\el)$ to each WSL. An estimation of this localization length $\xi_{\alpha}$ is obtained by computing the mean square spreading of WS states around their mean position $X^\alpha_n$:
\be
\begin{array}{ll}
\xi_{\alpha}^2&=\bra{\en^\alpha_n}(X-X_{n,\alpha})^2\ket{\en^\alpha_n}\\
&=\bra{\en^\alpha_n}(\ham_0-\en^0_{\alpha})^2\ket{\en^\alpha_n}/\el ^2,
\end{array}
\ee
where
\be
\en^{\alpha}_0(\el)\equiv \bra{\en^\alpha_n}\ham_0\ket{\en^\alpha_n}=\bra{\en^\alpha}\ham_0\ket{\en^\alpha},
\ee 
such that we can write
\be
\en^{\alpha}(\el)=\en^{\alpha}_0(\el)+\el X^{\alpha}.
\ee 
The quantity $\en^{\alpha}_0=\bra{\en^\alpha}\ham_0\ket{\en^\alpha}$ should not be mistaken for $\en^\alpha_{n=0}=\en^\alpha=\bra{\en^\alpha}\ham\ket{\en^\alpha}$ (they only agree at $\el=0$). We stress that for finite electric field $\el$, the WSL states $\ket{\en^\alpha_n}$ of a given ladder $\alpha$ are general linear combination that mixes Bloch states $\ket{\en^{\beta}(k)}$ of different band indices $\beta=1,...,N_b$ \cite{footnote4}. The above expression of the localization length shows that it can be estimated as $\xi_\alpha \sim W/\el$, where $W$ is the bandwidth. This agrees with the usual semi-classical expression for the size of Bloch oscillations, see e.g. \cite{Leo}.

Next we will focus on the thermodynamics of the unbounded Wannier-Stark spectrum and come back later to the exact expression of the energies $\en^\alpha(\el)$. For now, it is sufficient to know that they exist and depend on the electric field: the major difficulty of the statistical mechanics approach, namely the unbounded spectrum, is what we focus on next.

\subsection{Statistical mechanics of the Wannier-Stark ladder: local chemical potential}

The presence of an energy spectrum with no lower bound leads to an unstable ground state. In a large but finite system, it means that all the electrons are on one side of the crystal. Such a ground state is drastically different from the zero field ground state: the zero field ground state is translationally invariant and charge neutral at the scale of a unit cell. When turning on the field, the zero-field ground state happens to be a metastable state of the system: it is known that Zener tunneling from this state to the finite field ground state gives rise to a finite lifetime of the metastable state. However this lifetime happens to be large as the probability of Zener tunneling $\sim \exp(-\# \text{gap}^2/\el)$ is exponentially suppressed when the electric field goes to zero. From a statistical physics point of view, this means that the ergodicity time is much larger than the measurement time: the true ground state is not reached in practice and the system only explores states that are closely related to the zero-field ground state. On physical grounds, the polarization and susceptibility of the insulating crystal are related to how the zero-field ground state evolves into another translationally invariant and charge neutral (at the scale of the unit cell) state when the field is turned on. See for example, the discussion in \cite{Souza} and references therein.

Following this line, we ought to enforce translational invariance when we compute the free energy of the electrons in a small but finite electric field. Due to the structure of the spectrum, which is a set of WSL $\en_n^\alpha$ whose states within a ladder are related by the translation operator, enforcing translational invariance is simple: each rung $n$ (taking $N$ values) of a given ladder $\alpha$ (fixed) should be equally populated. While imposing this constraint is not an easy task when working with the free energy, the grand-canonical ensemble possesses a useful tool in order to do that: the chemical potential. This quantity is a Lagrange multiplier that enforces a specific average number of electrons. We now introduce one such Lagrange multiplier $\mu_n(\el)$ in each unit cell, tuned such that all the rungs of a single ladder are equally populated, and such that the total number of electrons still ensures the overall charge neutrality. In other words, we impose charge neutrality not only globally, but also \textit{locally}, in each unit cell. Within this approach the grand potential (\ref{gpot1}) becomes
\be
  \gpot(\{\mu_n\},\el,\beta) = -\frac{1}{\beta}\sum_{n,\alpha} \ln\l( 1 + e^{-\beta(\en_n^\alpha(\el) - \mu_n(\el))}\r)
\ee
and the constraint is realized for a local chemical potential
\be
  \mu_n(\el) =\el n + \tilde{\mu}(\el)= \mu^{(0)} + \el n + \el \mu^{(1)} + \mathcal{O}(\el^2)
\ee
where $\mu^{(0)}$ is the value of the chemical potential that realize the charge neutrality at zero field. Upon translation of all the energies, it can be set to $0$ (choice in the zero of energy). The term $\el n$ enforces that all rungs of a single ladder are populated equally, and $\mu^{(1)}$ is the first order correction to $\mu^{(0)}$. In appendix \ref{appendix1}, we show that
\be
  \mu^{(1)} = \frac{\sum_{\alpha} \l(\partial_\el \en^\alpha\r) n_F'(\overline{\en^\alpha})}{\sum_{\alpha} n_F'(\overline{\en^\alpha})}
\ee
where we have defined the average energy of the $\alpha$ Bloch band as $\overline{\en^\alpha} \equiv\int_{-\pi}^\pi \frac{dk}{2\pi}E^\alpha (k)$ (it is also the zero-field limit of the center of the $\alpha$ ladder $\en^\alpha(\el \to 0)$), $\partial_\el \en^\alpha \equiv \l.\partial_\el \en^\alpha(\el)\r\vert_{\el=0}$. Here we only derived the zeroth and first order in $\el$ of the chemical potentials $\mu_n(\el)$: the next orders are not needed if we are only interested in the polarization and the susceptibility as shown in the appendix \ref{appendix1}.

Once the constraint is imposed, the free energy of the electrons is
\be
  \free_{e^-}(N_{e^-},\el,\beta) \approx \sum_{n} q\mu_n  -\frac{1}{\beta} \sum_{n,\alpha}\ln\l( 1 + e^{-\beta(\en^\alpha(\el) - \el \mu^{(1)} )}\r)
\ee
which replaces (\ref{fee}) in the case of a local chemical potential. In the previous equation, we used that $\sum_n q =qN=N_{e^-}$. We then add the free energy of the ions (we recall that they are taken as static charges in a scalar potential), see equation (\ref{fions}), to obtain the total free energy of the system
\be
  \free(\el) \approx \sum_{n} q\el \mu^{(1)} -\frac{1}{\beta}\sum_{n,\alpha} \ln\l( 1 + e^{-\beta(\en^\alpha(\el) - \el \mu^{(1)} )}\r) 
\ee
Note that the contribution of ions cancels the $\el n$ term coming from $\mu_n$ in the total free energy $\free$.

We can now express the polarization (from its thermodynamic definition Eq. (\ref{thermopol})) as
\be
  \pol = - \sum_{\alpha} n_F\l(\overline{\en^\alpha}\r) \partial_\el\en^\alpha
  \label{pol}
\ee
as well as the susceptibility as
\ba
  \chi  & = 	& -\sum_\alpha \left[n_F\l(\overline{\en^\alpha}\r) \partial^2_\el \en^\alpha
		  +  n_F'\l(\overline{\en^\alpha}\r) \l(\partial_\el\en^\alpha\r)^2 \right] \nonumber \\
	& 	& + \l(\sum_{\alpha} n_F'(\overline{\en^\alpha}) \partial_\el \en^\alpha \r)^2 \l(\sum_{\alpha} n_F'(\overline{\en^\alpha})\r)^{-1}
  \label{susc}
\ea
Note that the above two formulas only require the knowledge of the energy spectrum (more precisely the center of the WSL $\en^\alpha(\el)$) at finite electric field in the limit of vanishing field. Eigenstates are not involved.

To summarize, the true ground state in the presence of a weak electric field is not reached during an experimentally accessible time due to exponentially suppressed Zener tunneling from the zero field ground state to the finite field ground state. We therefore made the assumption that instead of exploring the full phase space, the system in the presence of a weak electric field only explores the space of translationally invariant configurations (which are the configurations that are closest to the zero field ground state). Using this assumption, we derived the free energy and then obtained the polarization and susceptibility.

In appendix \ref{appendix2}, we explore a toy model in which we turn back to a single global chemical potential and add interactions between electrons in the form of an electrostatic cost for charge inhomogeneity. While the derivation is model specific, it shows that the polarization and the susceptibility obtained with a global chemical potential and for strong interactions agree with that obtained with a local chemical potential and no interactions. In other words, the main effect of electrostatic interactions is to enforce electro-neutrality within each unit cell.

\subsection{Perturbative-like expansion of the Wannier-Stark ladder energies}

The WSL are generated by the translation operator $T_a$. We call $|\en^\alpha_n\rangle$ a WS state belonging to the $\alpha^{th}$ WSL and with center position in the $n^{th}$ unit cell. This means that we can decompose the Hilbert space in orthogonal subspaces (labeled by $\alpha$) which are spanned by the families $\{\ket{\en^\alpha_n},n\}$, with $T_a\ket{\en^\alpha_n} = \ket{\en^\alpha_{n+1}}$ and $\ham\ket{\en^\alpha_n} = (\en^\alpha(\el) + \el n)\ket{\en^\alpha_n}$. Such families are stable under the translation operator, span subspaces that are orthogonal to one another and hence block-diagonalize the Hamiltonian. Reciprocally, if we find sufficiently many such subspaces (that is, as many subspaces as there are ladders in the finite field spectrum, or Bloch bands in the zero field spectrum) then each subspace is the subspace spanned by a single WSL: the Hamiltonian is block-diagonal and every block is part of a single ladder. Using the properties of the WSL spectrum, and taking a normalized $\ket{\psi^\alpha}$ that verifies $\ps{\psi^\alpha}{T_a\psi^\alpha} = 0$ (it does not necessarily need to be a WS state) in one of these subspaces, we have that the center of the $\alpha$ ladder is
\be
  \lim_{N\to\infty}\frac{1}{N}\sum_{n=-(N-1)/2}^{(N-1)/2} \psop{T_a^n \psi^\alpha}{\ham}{T_a^n \psi^\alpha} = \en^\alpha(\el)
\ee
where $N$, assumed to be odd, is the number of unit cells in the crystal. We assume here that the state $\ket{\psi^\alpha}$ is localized in the $n=0$ unit cell, i.e. $-\frac{1}{2} \leq \psop{\psi^\alpha}{X}{\psi^\alpha} < \frac{1}{2}$. If it is not the case, we apply the translation operator $T_a$ sufficiently many times to translate the state back to the $n=0$ unit cell.

To build a perturbative-like treatment, we use the Wannier states (or ``Wannier functions'') $\ket{w_n^\alpha}$ defined at zero electric field and  which constitute a basis of the Hilbert space. For isolated bands, they are defined as
\be
  \ket{w_n^\alpha} = \int_{BZ} \frac{\diff k}{\sqrt{2\pi}} e^{-ikn} \ket{\en^\alpha(k)}
\ee
where $\ket{\en^\alpha(k)}$ are the Bloch states for the band $\alpha$ of the zero-field Hamiltonian ($\ham_0\ket{\en^\alpha(k)}=\en^\alpha(k)\ket{\en^\alpha(k)}$). The Wannier functions have several interesting properties: (i) they block-diagonalize the zero-field Hamiltonian, and there are as many blocks as there are bands; (ii) for fixed $\alpha$, the family $\{\ket{w_n^\alpha},n\}$ is invariant under translation, i.e. $T_a\ket{w_n^\alpha} = \ket{w_{n+1}^\alpha}$; (iii) for suitable choices of the phase of the Bloch eigenvectors \cite{footnote3}, they are localized and as such, the matrix elements of the position operator are well-defined in the Wannier basis. Despite their name, the Wannier functions are not the WS states (they are not eigenstates of the Hamiltonian in the presence of an electric field). However their properties match those required by the presence of a WSL, hence we will use them as the starting point of our perturbative expansion. In a loose sense, Wannier functions $|w_n^\alpha\rangle$ are the $\el \to 0$ limit of WS states $|E_n^\alpha\rangle$.

We look for orthonormalized vectors $\ket{n,\alpha,\el}$ such that: (i) $\ket{n, \alpha,\el=0} \equiv \ket{w_n^\alpha}$ the Wannier functions; (ii) for any given value of the field, $T_a\ket{n,\alpha,\el} = \ket{n+1,\alpha,\el}$ so as to enforce the translational invariance of the family; and (iii) $\psop{m,\alpha,\el}{\ham}{n,\beta,\el} = 0$ for all $\alpha\neq\beta$ which ensures that the Hamiltonian is block-diagonalized. We do not require that the $\ket{n,\alpha,\el}$ are eigenstates of the Hamiltonian in the presence of a field (i.e. WS states), as this is not needed in order to recover the value of $\en^\alpha(\el)$.

Due the the translational invariance requirement, we can generically write the $\ket{n,\alpha,\el}$ as
\be
  \ket{n,\alpha,\el} = \ket{w^\alpha_n} + \el M_d(\el)^{\beta\alpha} \ket{w^\beta_{n+d}}
\ee
where a sum over repeated indices $\beta$ and $d$ is assumed. We can interpret the matrices $M_d(\el)$ as the Fourier coefficients of a periodic function $M(k,\el)$, where $k$ can be thought as a reciprocal vector in the first Brillouin zone (BZ). The states $\ket{n,\alpha,\el}$ need to be normalized and orthogonal to one another, and this conditions is given by
\be
  \l(\id + \el M(k,\el)\r)^\dagger\l(\id + \el M(k,\el)\r) = \id
\ee
and for $\alpha \neq \beta$, they must be orthogonal for the Hamiltonian, which is a condition expressed by
\ba
  \l(\id + \el M(k,\el)\r)^\dagger \ham(k) \l(\id + \el M(k,\el)\r) = 0  \nonumber \\
  \text{with } \ham(k) =  \l(\en(k) + e\el \bco(k) + e \el \frac{i}{2}\l(\overleftarrow{\partial_k} - \overrightarrow{\partial_k}\r)\r)
\ea
where $\en(k)$ is the matrix of the Bloch energies $\en^{\alpha\beta}(k) = \delta^{\alpha\beta}\en^{\alpha}(k)$, $\overleftarrow{\partial_k}$ (resp. $\overrightarrow{\partial_k}$) acts as a derivative on all the terms that are to its left (resp. right) and $\bco(k)$ is the matrix of Berry connection
\be
  \bco^{\alpha\beta}(k) = \im \ps{u^\alpha(k)}{\frac{\partial}{\partial k} u^\beta(k)}\, .
\ee
The cell-periodic Bloch state $\ket{u^\alpha(k)}$ (eigenstate of the zero-field Bloch Hamiltonian $\ham_0(k)=e^{-ikX}\ham_0 e^{ikX}$) is related to the Bloch eigenvector $\ket{\en^\alpha(k)}$ by
\be
  \ps{n,i}{\en^{\alpha}(k)} = e^{ik(n+x_i)}u^\alpha_i(k)
\ee
where $\ps{n,i}{\en^{\alpha}(k)}$ is the amplitude of the Bloch eigenvector on the site $i$ of the unit cell $n$. The position operator (see appendix A) is such that $X=\sum_{n,i}(n+x_i)\ket{n,i}\bra{n,i}$, where $n$ is the position of the unit cell ($n$ takes $N$ values) and $x_i$ is the position within the unit cell (or intra-cell position, with $i$ taking $N_b$ values).

These two constraints allow us to find $M(k,\el)$ order-by-order in the electric field, and the knowledge of $M(k,\el)$ allows us to take the trace on a single block of the Hamiltonian to get
\ba
  \en^\alpha(\el) &=& \int_{BZ}\frac{\diff k}{2\pi}\left( \en^{\alpha}(k) + \el \bco^{\alpha\alpha}(k) \right. \nonumber\\
		  & &- \left. \el^2 \sum_{\beta\neq \alpha } \frac{\bco^{\alpha\beta}(k)\bco^{\beta\alpha}(k)}{\en^{\beta}(k) - \en^{\alpha}(k)} + ...\right)\nonumber\\
		  &=&\overline{\en^\alpha}+\el\overline{\bco^{\alpha\alpha}}-\el^2 \sum_{\beta\neq \alpha } \int_{BZ}\frac{\bco^{\alpha\beta}\bco^{\beta\alpha}}{\en^\beta-\en^\alpha}
  \label{wslp}
\ea
We indicate an average over the BZ by $\overline{f}\equiv\int_{BZ}f \equiv \int_{BZ}\frac{\diff k}{2\pi}f(k)$ where $f(k)$ is any function of $k$.

In the zeroth order, one recognizes the mean value $\overline{\en^{\alpha}}$ of the energy of the $\alpha$-th Bloch band, a result already found in \cite{Wannier1960}. This is also the average energy of the $n=0$ Wannier state $\psop{w_0^\alpha}{\ham_0}{w_0^\alpha}$ in the absence of an electric field.

The first order term $\overline{\bco^{\alpha\alpha}}=\tfrac{Z^\alpha}{2\pi}$ is proportional to the Zak phase $Z^\alpha$ \cite{Zak1989} of the band and first appeared in \cite{Zak1968}. It is also related to the position of the $n=0$ Wannier state (a.k.a. the Wannier center) $\langle w_0^\alpha |X| w_0^\alpha\rangle=Z^\alpha/(2\pi)$. In other words the two first terms are simply the expectation value of the total energy in the Wannier state $\en^\alpha(\el)=\psop{w_0^\alpha}{(\ham_0+\el X)}{w_0^\alpha}+\mathcal{O}(\el^2)$. Although, this is strongly reminiscent of first order perturbation theory, below we argue that this is actually not the case.

These two first term of the WSL can also be obtained by the semiclassical quantization of Bloch oscillations, see for instance \cite{Resta2000,Niu2010}.

Surprisingly, the second order term in (\ref{wslp}) is not simply a second order perturbation formula like $\displaystyle\sum_{\beta\neq \alpha}\frac{\abs{\psop{w_n^\alpha}{X}{w_n^\beta}}^2}{\overline{\en^\alpha}-\overline{\en^\beta}}$ because the number and position of BZ integrals are not matching.

It is important to realize that the expansion of the WSL in powers of the field is not perturbative in the usual sense. Indeed, at zeroth order, the energy is the mean value of the energy of the Bloch band, which is not an eigenvalue of the Hamiltonian in absence of the electric field. Also, the WS states do not coincide with the Bloch eigenstates in the zero field limit. A crucial point is therefore to realize that in order to obtain the electric response of the crystal even in the low field limit, one has to use $\lim_{\el\to 0}\en_n^\alpha(\el)=\overline{\en^\alpha}$ instead of $\en^\alpha(k)$ as the energy spectrum suffers from a discontinuity at $\el=0$.

When choosing the Wannier functions $\ket{w^\alpha_n}$, we mentioned a phase (or gauge) choice: the Bloch eigenvectors $\ket{\en^\alpha(k)}$ may be multiplied by an arbitrary phase $e^{i\phi^\alpha(k)}$. Besides the fact that $e^{i\phi^\alpha(k)}$ has to be smooth and periodic over the Brillouin zone, there are no other restrictions. Indeed, if the aforementioned phase factor were not periodic or smooth, we would loose the localization properties of the Wannier functions. On the one hand, upon a gauge change, the off-diagonal Berry connection $\bco^{\alpha\beta}(k)$ is modified by the phase factor $e^{-i(\phi^\alpha(k)-\phi^\beta(k))}$, hence the product $\bco^{\alpha\beta}(k)\bco^{\beta\alpha}(k)$ is gauge invariant, and so is the second order of the WSL energies. On the other hand, the diagonal Berry connection $\bco^{\alpha\alpha}(k)$ is modified by the total derivative $\partial_k \phi^\alpha$ whose integral over the Brillouin zone is quantized to an integer (which counts how many times the phase winds around the origin). But remember that we have previously required that the vector $\ket{n=0, \alpha,\el}$ is located in the $n=0$ unit cell, which in turn imposes that the Wannier function $\ket{w_0^\alpha}$ has its center in the zeroth unit cell. Transforming to a gauge where  $e^{i\phi^\alpha(k)}$ winds one extra time around the origin amounts to translation by one unit cell all the Wannier functions of the band $\alpha$. The spectrum being unbounded by both above and below, an unambiguous definition of the WSL imposes that $\ket{w_0^\alpha}$ must be situated in the zeroth unit cell. Translating it back, amounts to effectively cancel the extra winding of the phase. The above expression is hence gauge invariant. It is actually well-known that the Zak phase is gauge-invariant despite its being an open-path geometric phase; see, for example, the nice discussion in Ref. \cite{Resta2000}. Note, however, that the Zak phase depends on the choice of position origin. Here, we have made the choice that the charge weighted barycenter of the ions $\bar{x}=0$ in the $n=0$ unit cell.

\subsection{Full expression of the polarization and the susceptibility}
Before giving the full expressions of the polarization and of the susceptibility -- i.e. essentially inserting (\ref{wslp}) in (\ref{pol}) and (\ref{susc}) --, we recall the hypotheses we have used in their derivation: (i) we restrict to uniform filling of the WSL states, which is a valid approximation at low electric field (suppressed Zener tunneling) and low temperature (both with respect to the gap and to the electrostatic interaction energy, i.e. costly charge inhomogeneities); (ii) the origin of position is taken as the charge-weighted barycenter of the ions in the $n=0$ unit cell; (iii) the phases of the Bloch eigenvectors are such that the Wannier functions $\ket{w_0^\alpha}$ are localized in the zeroth unit cell.

With these hypotheses, using the perturbative expression of the WSL energies found in the previous section and restoring all constants that were previously set to 1, we reach
\be
\pol = - \frac{e}{a} \sum_\alpha n_F\l(\overline{\en^\alpha}\r) \overline{\bco^{\alpha\alpha}}=-e \sum_\alpha n_F\l(\overline{\en^\alpha}\r) \frac{Z^\alpha}{2\pi}
\label{polfinal}
\ee
and
\ba
\chi & = & \frac{e^2}{a} \sum_{\alpha,\beta\neq \alpha} n_F\l(\overline{\en^\alpha}\r)\int_{BZ} \frac{\bco^{\alpha\beta}\bco^{\beta\alpha}}{\en^\beta-\en^\alpha}\label{suscfinal} \\
     &-& \frac{e^2}{a}\sum_\alpha n_F'\l(\overline{\en^\alpha}\r)\overline{\bco^{\alpha\alpha}}^2 
     + \frac{e^2}{a}\frac{[\sum_\alpha n_F'\l(\overline{\en^\alpha}\r)\overline{\bco^{\alpha\alpha}} ]^2}{\sum_\alpha n_F'\l(\overline{\en^\alpha}\r)}\nonumber
\ea
where $\bar{f}\equiv a\int_{BZ}\frac{dk}{2\pi} f(k)$. At zero temperature, we recover the well-known formula of King-Smith, Vanderbilt and Resta \cite{KSV,VanderbiltResta} for the polarization
\be
\pol = -\frac{e}{a}\sum_{\alpha\ occ.}\overline{\bco^{\alpha\alpha}}= -e\sum_{\alpha\ occ.}\frac{Z^\alpha}{2\pi}
\ee
and a recent result of Swiecicki and Sipe \cite{SwiecickiSipe} for the susceptibility
\be
  \chi = \frac{e^2}{a} \sum_{\alpha\ occ.} \sum_{\beta\neq\alpha} \int_{BZ} \frac{\bco^{\alpha\beta}\bco^{\beta\alpha}}{\en^\beta-\en^\alpha}\geq 0
\ee
The susceptibility is positive, in agreement with a general argument \cite{LandauLifshitz8}. In the above formula, the sum over $\alpha$ is restricted to occupied bands.

\subsection{Quantum of polarization}
At zero temperature, the electric polarization of a bulk crystal is defined up to a \emph{quantum of polarization}, which is an integer in the proper units \cite{VanderbiltResta}. The \emph{quantum of polarization} means that from the bulk point of view, the polarization cannot be defined in an absolute manner: as long as the surface of the crystal is not specified, we can only get the difference of polarization between two configurations of the crystal. For instance one can access unambiguously the change of polarization upon a change of the applied stress by only looking at the bulk. Then an adiabatic pumping argument shows that two identical configurations in the bulk can have a difference of polarization which is an integer. Hence an absolute value of the bulk polarization has to be defined up to an integer. 

In the above formula for the polarization (\ref{polfinal}), the quantities which are defined up to an integer are the Wannier centers $\overline{\bco^{\alpha\alpha}}$, meaning that the finite temperature formula we give does not obviously possess this \emph{quantum of polarization}. To recover it, we need to recall that the spectrum is made of several WSL of the form $\en^\alpha(\el) + \el n$, and that each rung correspond to a localized eigenstate. To unambiguously define the different energies $\en^\alpha(\el)$, we have imposed that the $n=0$ eigenstates of the different ladders belong to the same unit cell: it would make no sense to compare the energy of a state that is located in the $m^{th}$ unit cell to the energy of one other located in the $n^{th}$ unit cell, as the latter would feel an extra electric potential $\el(m-n)$, hence have its energy shifted by $\el(m-n)$ with respect to the former eigenstate. Now, to change the value of the Wannier center $\overline{\bco^{\alpha\alpha}}$ by one, we need to make a gauge choice in which the phase of the Bloch eigenvectors winds an extra time around the origin when we go from one side of the Brillouin zone the other. But this extra winding amounts to move the WS states of the ladder $\alpha$ by one unit cell, which we cannot do unless we also move the other ladders, as we would then compare the energies of the different ladders by comparing the energy of states in different unit cells.

So if we change the Wannier center of one band $\overline{\bco^{\alpha\alpha}}$ by the integer $p$, then we must change it for all the bands at once, and the change of polarization we get would then be
\be
\sum_\alpha p\, n_F(\overline{\en^\alpha}) = p
\ee
and we therefore recover the \emph{quantum of polarization} also at finite temperature. This fact also lead to the gauge invariance of the susceptibility $\chi$ at finite temperature. Indeed, the quantity
\be
  - \sum_\alpha n_F'\l(\overline{\en^\alpha}\r)\overline{\bco^{\alpha\alpha}}^2
  + \frac{\l(\sum_\alpha n_F'\l(\overline{\en^\alpha}\r)\overline{\bco^{\alpha\alpha}} \r)^2}{\sum_\alpha n_F'\l(\overline{\en^\alpha}\r)}
\ee
does not change when we shift simultaneously the Wannier centers $\overline{\bco^{\alpha\alpha}}$.

Along with the presence of a quantum of polarization, the polarization and susceptibility should be invariant both under a change of the origin of position and under a change of the unit cell. The former invariance is a direct consequence of charge neutrality and is easily checked. The latter is harder to verify because the Berry connection does not trivially change under a change of the unit cell. We did check it for every example we considered, however.



\section{Toy models: analytics versus numerics}
To check our analytical predictions, we now consider two toy-models that can be solved exactly either analytically or numerically.

\subsection{Chain of uncoupled dimers}
The first model is an infinite chain of uncoupled dimers, i.e. a chain of molecules made of two different atoms $A$ and $B$, each with a single orbital. Atoms are located at $x_A+n$ and $x_B+n$, where $n$ is an integer (we set the lattice spacing $a=1$). Each dimer is characterized by an intra-dimer hopping amplitude $t=1$ and on-site energies $\pm \Delta$ for the two sites forming the dimer. There are no inter-dimer hopping amplitudes, which greatly simplifies the problem. In this case, it is obvious that the electric response of the crystal is identical to that of a single dimer, which is easily computed. For the n$^{th}$ dimer (and taking the mean ion position in the $n=0$ unit cell as the origin $(x_A+x_B)/2=0$) the Hamiltonian in an electric field reads:
\be
\ham_\el=\left(\begin{array}{cc}\Delta + \frac{x_A-x_B}{2}\el & 1 \\ 1& - \Delta - \frac{x_A-x_B}{2}\el\end{array}\right) +n \el
\ee
The model depends on two parameters ($\Delta$ and $x_A-x_B$) and on the applied electric field $\el$. The eigen-energies are
\be
\en_n^\pm (\el)=\en^\pm (\el)+n\el=\pm \sqrt{(\Delta+\frac{x_A-x_B}{2}\el)^2+1}+n\el
\ee
which are indeed two WSL labeled by $\alpha=\pm$. Expanding to second order in the electric field, we find that the WSL centers are
\be
\en^\alpha (\el)\approx \alpha \sqrt{\Delta^2+1} +\alpha \frac{(x_A-x_B)\Delta}{2\sqrt{\Delta^2+1}}\el +\alpha \frac{(x_A-x_B)^2}{8(\Delta^2+1)^{3/2}}\el^2
\label{wsldimers}
\ee
This should be compared to the perturbative-like result given in equation (\ref{wslp}) and involving the dispersion relation and the diagonal and off-diagonal Berry connections. In order to compute the latter, we need the zero-field Bloch Hamiltonian:
\be
  \ham_0(k)=e^{-ikX}\ham_0e^{ikX}=\left(\begin{array}{cc}\Delta  & e^{-ik(x_A-x_B)} \\ e^{ik(x_A-x_B)}& - \Delta\end{array}\right)
\ee
The energy bands have a flat dispersion relation $\en^\alpha(k)=\alpha\sqrt{\Delta^2+1}$ and therefore $\overline{\en^{\alpha}}=\alpha \sqrt{\Delta^2+1}$ which matches the zeroth order in the WSL ladder (\ref{wsldimers}). The cell-periodic part of the Bloch states are
\ba
  \ket{u^+(k)} & = & \l(\cos \tfrac{\theta}{2}e^{-i\tfrac{\phi}{2}}, \sin\tfrac{\theta}{2}e^{i\tfrac{\phi}{2}}\r) \nonumber \\
  \ket{u^-(k)} & = & \l(-\sin \tfrac{\theta}{2}e^{-i\tfrac{\phi}{2}}, \sin\tfrac{\theta}{2}e^{i\tfrac{\phi}{2}}\r)
\ea
[in the periodic gauge where $\ket{\en^\alpha(k+2\pi)} = \ket{\en^\alpha(k)} \Rightarrow u_i^\alpha(k+2\pi) = e^{-2i\pi x_i}u_i^\alpha(k)$], and
\be
\cos \theta = \tfrac{\Delta}{\sqrt{\Delta^2+1}},\ \sin \theta = \tfrac{1}{\sqrt{\Delta^2+1}},\ \phi=k(x_A-x_B)
\ee
The diagonal Berry connection is also independent of $k$
\be
  \overline{\bco^{\alpha\alpha}} = \bco^{\alpha \alpha} = \alpha \tfrac{1}{2}(\partial_k \phi)\cos\theta = \alpha \tfrac{1}{2}(x_A-x_B) \tfrac{\Delta}{\sqrt{\Delta^2+1}}
\ee
and we recognize the first order of the WSL ladder of (\ref{wsldimers}). Finally, the off-diagonal Berry connection is
\be
  \bco^{-+}=\bco^{+-}=-\tfrac{1}{2}(\partial_k \phi)\sin\theta
\ee
so that the second order of the perturbative expansion is
\be
  \sum_{\beta\neq \alpha}\frac{\bco^{\alpha\beta}\bco^{\beta\alpha}}{\en^\alpha(k)-\en^\beta(k)} = \alpha\frac{(x_A-x_B)^2}{8(\Delta^2+1)^{3/2}}
\ee
recovering the second order of equation (\ref{wsldimers}). The perturbative-like expansion of the WSL energies is therefore correct for the chain of dimers. Thus it can be safely used in the thermodynamic derivation of the electric polarization and susceptibility. Note also that in the case of a chain of decoupled dimers, the use of a local chemical potential is clearly justified as each dimer is independently half-filled even in the presence of an electric field.

\subsection{Rice-Mele chain}
In order to study solitons in polymer chains such as polyacetylene, Rice and Mele proposed a tight-binding model of a dimerized chain with staggered on-site potential \cite{RiceMele1982}. It is a standard toy-model in the study of the electric polarization of crystals \cite{KSV,Niu2010}. 


The chain is made of an alternating succession of sites $A$ and $B$ occupied by cations and carrying each half an electron charge $e/2$ (this is related to considering spinless electrons). The sites are equally spaced so that $x_A-x_B=\tfrac{1}{2} + n$, where $n$ is an integer. The Bloch Hamiltonian is given by
\be
  \ham_0(k) = \l(
    \begin{array}{cc}
      \Delta  						& 2t(\cos \frac{k}{2} - i\delta \sin\frac{k}{2}) \\
      2t(\cos \frac{k}{2} + i\delta \sin\frac{k}{2})	& - \Delta
    \end{array}\r)
\ee
where $\pm \Delta$ are the on-site energies on the two sublattices and the two hopping amplitudes are $t(1\pm \delta)$. In the following we set $t=1$ in addition to $a=1$ and $e=1$. The chain of uncoupled dimers studied in the previous section corresponds to $\delta = 1$ and $t = \frac{1}{2}$ while $\Delta\neq0$ and $x_A-x_B$ should not be restricted to $\tfrac{1}{2} + n$. The energy spectrum at zero electric field is
\be
  \en^{\pm} (k) =\pm\sqrt{\Delta^2+4\cos^2\frac{k}{2}+4\delta^2\sin^2 \frac{k}{2}}
\ee

For simplicity and following \cite{KSV}, we set $\Delta=\Delta_0 \cos\theta$ and $\delta=\delta_0 \sin \theta$ and use the single angular parameter $\theta$ to tune the model by choosing $\Delta_0=\delta_0=0.6$ as an example. 

\subsubsection{WSL: numerics on finite versus analytics for infinite chain}
\begin{figure}
  \psfrag{stddev}[cc][cc][1][180]{Standard deviation}
  \psfrag{efieldstddev}[cc][cc][1]{Electric field}
  \psfrag{energy}[cc][cc][0.6][180]{Energy}
  \psfrag{efielddspectrum}[cc][cc][0.6]{Electric field}
  \includegraphics[width=8cm]{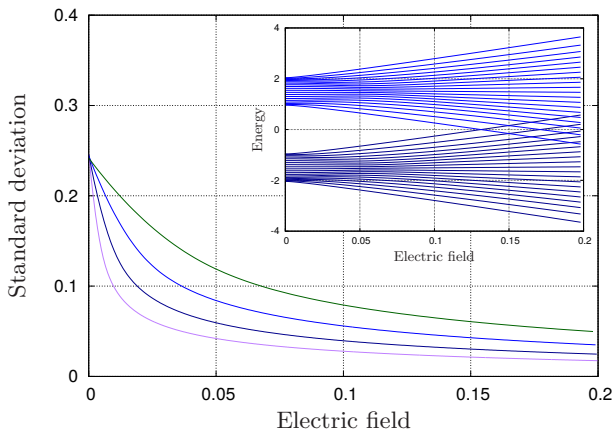}
  \caption{Inset: Spectrum of the Rice-Mele chain for $\theta = \tfrac{3}{4}\pi$, in the presence of an electric field. For readability, the spectrum correspond to a small chain of 20 unit cells. At zero field, the chain is a two band insulator, while at finite field, the bands evolve into a WSL with energies $\en^{\pm}_n(\el) = \overline{\en^\pm} + \el n +  \el\overline{\bco^{\pm\pm}} + ...$, see Eq. (\ref{wslp}). Main panel: Standard deviation of the numerical energy levels w.r.t. the analytical levels of an infinite chain (up to second order, see Eq. (\ref{wslp})), i.e. $\sqrt{\tfrac{1}{N}\sum_n \l(\en_{n,\text{numeric}}^- - \en_{n,\text{analytic}}^-\r)^2}$, for $N = 20$ (highest deviation), 40, 80 and 160 (lowest deviation) unit cells. The initial decrease of the deviation is exponential $e^{-\el/\el_c}$ and characterized by the field $\el_c^{}\sim \frac{W}{N}$ where $W$ is the bandwidth. High-field decrease of the deviation w.r.t to the Wannier-Stark ladder of the infinite system is governed by a second characteristic value of the field. Units are such that $e=1$, $\hbar=1$ and $a=1$.}
\label{fig:RiceMele-WSL}
\end{figure}

In the case of the Rice-Mele chain, in contrast to the dimer chain, it is not possible to analytically obtain the energy spectrum of an infinite chain in the presence of an electric field. However, we can numerically obtain the spectrum for a finite chain with an electric field and compare it with equation (\ref{wslp}), which gives the perturbative-like expansion in powers of the electric field in the thermodynamic limit, see Figure \ref{fig:RiceMele-WSL}. The agreement becomes very good when the electric field is sufficiently large that finite size effects are negligible (i.e. $\el \gg \frac{W}{N}$ where $W$ is the bandwidth) and sufficiently small to be in the weak field regime (i.e. $\el \ll W$) and also that the order $\el^2$ expansion of the WSL is valid (corresponding to an even larger electric field). These inequalities are equivalent to requiring that the WS localization length $\xi \sim \frac{W}{\el}$ be smaller than the system size $Na=N$ and larger than the lattice spacing $a=1$. In summary, the WSL regime of a bulk crystal exists in a finite system provided that $\el \gg \frac{W}{N}$. In addition, one explores the weak field limit provided that $\el \ll W$. Figure \ref{fig:RiceMele-WSL} also shows that the first level crossing between levels coming from different bands occurs at an electric field $\sim \frac{\text{gap}}{N}$. This is of the similar to $\frac{W}{N}$ as the gap and the bandwidth are taken to be of the same order.

\begin{figure}
  \psfrag{ecenter}[cc][cc][1][180]{$\en_n^\alpha - \el X_n^\alpha$ (Energy center)}
  \psfrag{position}[cc][cc][1]{$X_n^\alpha$ (Position of the eigenstate)}
  \psfrag{hfield}[cr][cr][0.6]{Higher electric field}
  \psfrag{lfield}[cr][cr][0.6]{Lower electric field}
  \includegraphics[width=8cm]{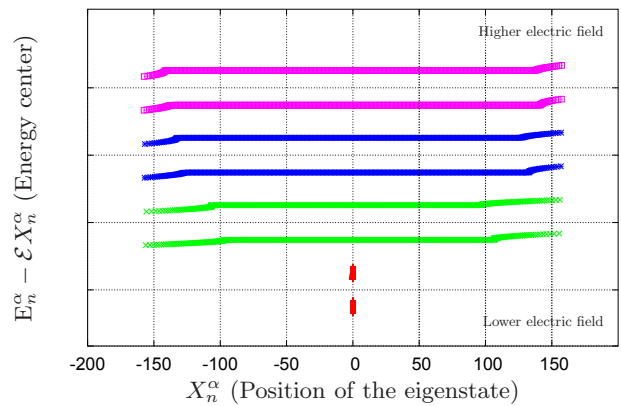}
  \caption{Energy center $\en_n^\alpha - \el X_n^\alpha$ as a function of the average position $X_n^\alpha$ for eigenstates of a finite chain with $N = 320$ unit cells) in an electric field. $n$ is the unit cell index and $\alpha$ is the index labeling the two bands. At weak electric field, $\el \ll W/N$, the energy center spans the zero-field bandwidth and the average position is the chain center for all eigenstates (the two bands are visible in red in the figure). When increasing the electric field and once the WSL regime is reached (green), bulk eigenstates form a plateau and all have the same energy center. Edge effects are seen on the two ends of the chain and tend to disappear with increasing field (blue and magenta). The curves in different colors are shifted vertically for clarity; the typical (vertical) distance between curves of the same color is of the order of the band gap.}
\label{fig:plateau}
\end{figure}

A convenient way of identifying this WSL regime is to plot the ``energy center'' $\en_n^\alpha - \el X_n^\alpha$, where $\en_n^\alpha(\el)$ is the energy of a numerically obtained eigenstate and $X_n^\alpha(\el)$ is its average position, as a function of $X_n^\alpha$ for a given band $\alpha$ (see Figure \ref{fig:plateau}.). Indeed, in the WSL regime, the energy spectrum should be given by equation (\ref{wslp}), which shows that $\en_n^\alpha - \el X_n^\alpha\approx \overline{\en^\alpha}+\mathcal{O}(\el^2)$ is almost field-independent. When the electric field is smaller than $\frac{W}{N}$ and negligible, almost all eigenstates have the same average position at the center of the chain and eigen-energies that vary continuously between the bottom and the top of the zero-field band (see the red points in Figure \ref{fig:plateau}). Then, when the field becomes larger than $\sim \frac{W}{N}$, eigenstates become localized in different unit cells ($X_{n+1}^\alpha-X_n^\alpha\approx 1$), but all have the same $\en_n^\alpha - \el X_n^\alpha$ forming a plateau as a function of the average position (see the green curve). The plateau is electric field independent and given by $\en_n^\alpha - \el X_n^\alpha\approx \overline{\en^\alpha}$ until the electric field becomes larger than $\sim W$. Then the plateau starts to depend on the electric field in a quadratic manner $\en_n^\alpha - \el X_n^\alpha\approx \overline{\en^\alpha}+\mathcal{O}(\el^2)$ revealing the electric susceptibility. The only deviations from this typical behavior are found near the edges of the finite chain.



\subsubsection{Polarization and susceptibility}
\begin{figure}
  \psfrag{xlabel}[ct][cc]{$\theta\ (\Delta = 0.6\cos\theta,\ \delta = 0.6\sin\theta)$}
  \psfrag{ylabel}[cc][cc]{$-\partial_\el \en^{-}$}
  \psfrag{p2}[cc][cc]{$\tfrac{\pi}{2}$}
  \psfrag{p}[cc][cc]{$\pi$}
  \psfrag{3p2}[cc][cc]{$\tfrac{3\pi}{2}$}
  \psfrag{2p}[cc][cc]{$2\pi$}
  \includegraphics[width=8cm]{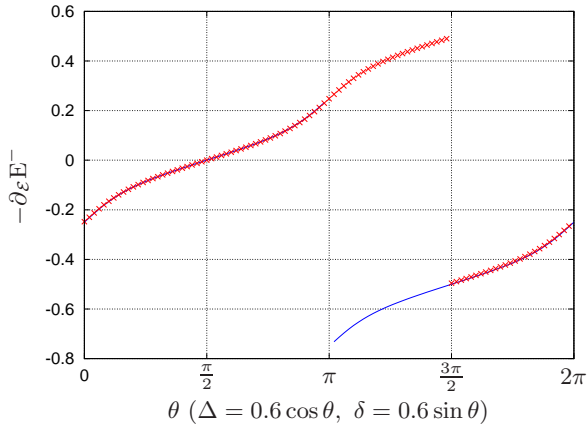}
  \caption{Zero-temperature spontaneous polarization $\pol$ [in units such that $e=1$] as a function of the Rice-Mele parameter $\theta$: $\theta=0$ and $\pi$ correspond to a CDW chain and $\theta=\pi/2$ and $3\pi/2$ to an SSH chain. The numerically computed first order of the WSL $-\partial_\el \en^-|_{\el=0}$ emerging from the valence band of a finite Rice-Mele chain with 80 unit cells is shown with red crosses. The analytical prediction of equation (\ref{wslp}) for the infinite system $\pol=-X^-$ is shown as a blue line, where $X^-$ is the Wannier center of the lower band [in units such that $a=1$]. For the infinite system (blue line), the polarization is defined modulo the quantum of polarization, which is 1 here, such that $-1/2\leq \pol <1/2$. For the finite chain, the polarization has an absolute meaning and is not defined modulo a quantum of polarization. When $\pi<\theta<2\pi$, the finite chain has two edge states with opposite energies inside the bulk gap.}  
  \label{fig:RiceMele-WSLFirstOrder}
\end{figure}

The first order term in $\el$ of the trace of $\ham_\el = \ham_0 + \el X$ on the WSL emerging from the lower band is presented in Figure \ref{fig:RiceMele-WSLFirstOrder} (see the red crosses) as a function of the parameter $\theta$ for a finite Rice-Mele chain with 80 sites. This is essentially the zero temperature polarization. It is compared with the Wannier center (or Zak phase divided by $2\pi$) for the lower band computed for the infinite system (see the blue full line).

For the infinite system, the polarization is defined modulo 1 and $\pol\to -\pol$ under inversion. Inversion symmetry is only present at particular values of $\theta$, implying that $\pol=-\pol$ modulo 1. These remarkable values of the parameters are: $\theta=0$ or $\pi$ corresponding to a charge density wave (CDW) like chain, with site-centered inversion symmetry resulting in a quantized spontaneous polarization $\pol = \pm \frac{1}{2}$; and $\theta=\tfrac{\pi}{2}$ or $\tfrac{3\pi}{2}$ corresponding to a Su-Schrieffer-Heeger (SSH) chain \cite{SSH}, with bond-centered inversion symmetry leading to a vanishing spontaneous polarization $\pol = 0$. Note that from a bulk perspective, the two SSH phases $\theta=\tfrac{\pi}{2}$ and $\tfrac{3\pi}{2}$ are identical and cannot be distinguished. Their difference of behavior is only revealed upon introducing an edge. In particular, the bulk polarization cannot be used to characterize the SSH as a 1D topological insulator as it vanishes in both phases \cite{ChenLee2011a}.

However, for a finite chain, the polarization can be given an absolute meaning (i.e. without the modulo inherent to the quantum of polarization) because once the edges are specified, the polarization becomes a well-defined quantity. In the $\theta \in [\pi,2\pi]$ range, the chain with an even number of sites possesses one localized state at each end of the chain with opposite energies within the bulk gap. The jump in polarization at $\theta=3\pi/2$ happens when both edge states cross zero energy. For the finite chain, there is now a clear difference in polarization between the SSH chain at $\theta=\pi/2$ for which the polarization vanishes and $\theta=3\pi/2$ for which the polarization jumps from $1$ to $-1$. The first phase is considered to be trivial and the second to be topological.

\begin{figure}
  \psfrag{xlabel}[tc][cc]{$\theta\ (\Delta = 0.6\cos\theta,\ \delta = 0.6\sin\theta)$}
  \psfrag{ylabel}[cc][cc]{$-2\partial_\el^2 \en^{-}$}
  \psfrag{p2}[cc][cc]{$\tfrac{\pi}{2}$}
  \psfrag{p}[cc][cc]{$\pi$}
  \psfrag{3p2}[cc][cc]{$\tfrac{3\pi}{2}$}
  \psfrag{2p}[cc][cc]{$2\pi$}
 \includegraphics[width=8cm]{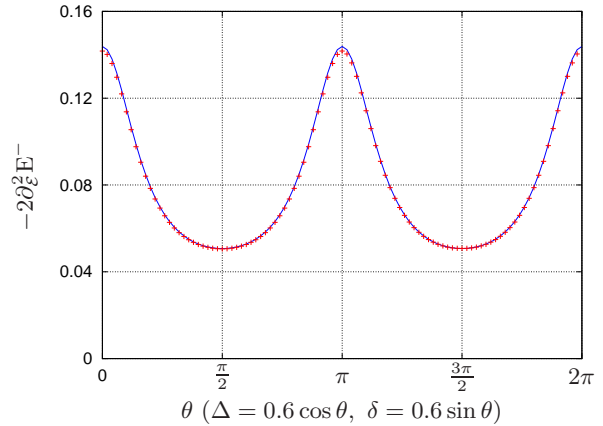}
\caption{Numerically computed second order of the energy of the WSL $-2\partial^2_\el \en^-|_{\el=0}$ emerging from the lowest band of the Rice-Mele model as a function of the angular parameter $\theta$ (red crosses) compared with the analytical value of equation (\ref{wslp}) (blue line). This is essentially the susceptibility $\chi$ at zero temperature. The calculation was done using 80 unit cells; finite size effects explain the small difference between the numerical and analytical curves around $\theta = 0, \pi$. Units are such that $e=1$, $\hbar=1$ and $a=1$.
\label{fig:RiceMele-WSLSecondOrder}}
\end{figure}

Figure \ref{fig:RiceMele-WSLSecondOrder} presents the second order term in $\el$ of the trace of $\ham_\el = \ham_0 + \el X$ on the WSL emerging from the lower band. This is essentially the zero temperature susceptibility. Small finite size effects can be noted at the second order around $\theta=0$ and $\theta=\pi$. This behavior of the susceptibility as a function of $\theta$ qualitatively follows that of the square of the localization length of the maximally localized Wannier state.

Breaking particle-hole symmetry by adding a term $\en_0(k)\sigma_0$ (where $\sigma_0$ is the $2\times 2$ identity matrix) to the Hamiltonian of the Rice-Mele model changes the energy spectrum but not the wavefunctions. Therefore it does not change the polarization and the susceptibility as the WSL -- i.e. the energy levels given in equation (31) -- are left unchanged.

\section{Conclusion}
In the present work, we have devised a statistical mechanics approach to the electric properties -- polarization and susceptibility -- of a band insulator at finite temperature. The key steps consist in, first, computing the Wannier-Stark ladder energy spectrum in a perturbative-like fashion at second order in the electric field and, second, in obtaining a relevant thermodynamical potential by imposing a local electroneutrality within each unit cell in the presence of the electric field. Our main results are equation (\ref{wslp}) for the WSL energy spectrum at second order in the electric field, equation (\ref{polfinal}) for the electric polarization at finite temperature and equation (\ref{suscfinal}) for the electric susceptibility at finite temperature. The correctness of the perturbative WSL energy spectrum (\ref{wslp}) was checked by comparing it with exact calculations in the case of two different toy-models (a chain of uncoupled dimers and a Rice-Mele chain). One advantage is that the same method can be used to compute response functions at first and second order (and actually also at higher orders).

Qualitatively, at zero temperature and in the simplest two-band model, the spontaneous polarization is essentially given by the Wannier center, i.e. the average position of the Wannier function (which is gauge independent), as found by King-Smith, Vanderbilt and Resta using a quite different approach. For the susceptibility, the physical interpretation is more complicated. At zero temperature, it is qualitatively given by the square of the localization length of the maximally localized Wannier function \cite{footnote5,MarzariRMP2012}, divided by an energy gap which is the energy difference between the average energies of the two bands. Indeed
\be
\chi = \sum_{\alpha\ occ.} \sum_{\beta\neq\alpha} \int_{BZ} \frac{\bco^{\alpha\beta}\bco^{\beta\alpha}}{\en^\beta-\en^\alpha}\sim \frac{\langle w_n^-|(\Delta X)^2|w_n^- \rangle}{\overline{\en^+}-\overline{\en^-}}
\ee
where $\sqrt{\langle w_n^-|(\Delta X)^2|w_n^- \rangle}$ is the localization length of the maximally localized Wannier function of the band $\alpha=-$ and $\Delta X=X-\langle X \rangle$.

Eventually, we mention the difficulty of using a gauge-invariant perturbative approach based on Green's functions to compute the density of states in the presence of an electric field (the polarization is related to the first derivative of the density of states w.r.t. the field and the susceptibility to the second derivative). Such an approach was, for example, proposed in \cite{ChenLee2011b} for both the electric and magnetic responses. Whereas it works well for the magnetic field, allowing to compute the magnetization and the orbital magnetic susceptibility \cite{Raoux2015}, it encounters severe difficulties in the case of an electric field. In particular, \cite{ChenLee2011b} have to assume that the finite-field polarization is given by the Zak phase in order to use their method but can not derive this fundamental relation.

Apart from the present approach, we are aware of two others that amount to imposing a local electronic filling within each unit cell. First, one can directly replace the linearly rising electric potential in the Hamiltonian by a piecewise linear potential (such as sawtooth or continuous triangular) with either the periodicity of the Bravais lattice or a supercell periodicity (see for example \cite{SpringborgKirtman} for a discussion). Then the energy spectrum remains that of a periodic system and usual thermodynamics can be employed automatically resulting in electro-neutrality within each cell. The drawback of this approach is that it does not recover the correct Zak phase formula for the electric polarization, although it has been used to compute higher order responses such as the electric susceptibility. A second approach would consist in defining a local density of states (involving not only the WSL energy spectrum but also the WS states) in order to impose the local electronic filling by a local chemical potential. We tried this approach -- which we find physically quite appealing -- and were surprised to realize that it also does not recover the KVR formula for the electric polarization \cite{fredpunpublished}.



\acknowledgements
We acknowledge useful discussions with Gilles Montambaux and Lih-King Lim especially on the Zak phase.

\appendix

\section{Position operator}
\label{positionoperator}
In this appendix, we discuss more precisely the position operator and its action on the WS states. The position operator 
\be
X=\sum_{n} \sum_{i=1}^{N_b} X_i \ket{n,i}\bra{n,i}
\ee
can be split into two distinct parts $X=x+R$. The contribution $x$ -- the intra-cell position operator -- is defined by
\be
x=\sum_n \sum_{i=1}^{N_b} x_i \ket{n,i}\bra{n,i},
\label{xcell}
\ee
and is translationally invariant $x=T_a^{-n} x T_a^n$. As a consequence we can write
\be
x^{\alpha}=\bra{\en^\alpha_n}x\ket{\en^\alpha_n}=\bra{\en^\alpha}x\ket{\en^\alpha},
\ee
By contrast, the contribution $R$ -- the Bravais lattice position operator -- is defined by
\be
R=\sum_n \sum_{i=1}^{N_b} n\ket{n,i}\bra{n,i}.
\label{xbravais}
\ee
and verifies the ladder identity $T_a^{-n} R T_a^n=R+n\id$. Note that in each unit cell, it is simply proportional to the identity. For this contribution we can write
\be
R^\alpha=\bra{\en^\alpha_n}R\ket{\en^\alpha_n}=\bra{\en^\alpha}R\ket{\en^\alpha}+n=r^{\alpha}+n,
\ee
For each ladder, the previously defined position center is thus the sum of two distinct contributions $X^{\alpha}=x^{\alpha}+r^{\alpha}$. On the one side the contribution $x^{\alpha}$ measures the intracell asymmetry of the probability of WS states $\ket{\en^\alpha_n}$; on the other side the contribution $r^{\alpha}$ measures the intercell asymmetry of the probability of WS states $\ket{\en^\alpha_n}$. 

\section{Chemical potential as a function of the field}
\label{appendix1}
In this appendix, we show that we only need the dependence of the chemical potential on the electric field at first order in order to obtain the susceptibility. The total free energy is
\be
  \free = \sum_{n,\alpha} \mu_n -\frac{1}{\beta} \sum_{n,\alpha} \ln(1+e^{-\beta(\en_n^\alpha(\el) - \mu_n)}) -\sum_n q\el n
\ee
where $\en_n^\alpha(\el)=\en^\alpha(\el)+\el n$ and in the local chemical potential approach, $\mu_n=\tilde{\mu}(\el)+\el n$, with $\tilde{\mu}(\el)=\mu^{(0)}+\el \mu^{(1)}+...$. The free energy per unit cell is therefore
\be
  \frac{\free}{N} = q\tilde{\mu}(\el) -\frac{1}{\beta} \sum_{\alpha} \ln(1+e^{-\beta(\en^\alpha(\el) - \tilde{\mu}(\el))})
\ee
Taking a derivative with respect to the field, we find that the polarization at finite electric field is
\be
  \pol(\el)=-\sum_{\alpha}\partial_\el \en^\alpha(\el) n_F(\en^\alpha(\el)-\tilde{\mu}(\el))
\ee
We used that the number of electron in each unit cell is fixed by the requirement of local electro-neutrality so that 
\be
q=\sum_\alpha n_F(\en^\alpha(\el)-\tilde{\mu}(\el)) \, .
\label{qen}
\ee 
Taking a second derivative with respect to the field, we find that the polarizability is
\be
  \chi=\sum_\alpha \left[n_F(\en^\alpha) \partial^2_\el \en^\alpha  + n_F'(\en^\alpha) \partial_\el \en^\alpha (\partial_\el \en^\alpha - \partial_\el \tilde{\mu}) \right]
\ee
At zero electric field, only $\tilde\mu \to \mu^{(0)}$ and $\partial_\el \tilde\mu \to \mu^{(1)}$ appear in the expression of the polarization and susceptibility.
As $\mu^{(0)}$ can conveniently be set to $0$ (by a choice of the origin of energy), we only need to know the first derivative of the chemical potential with respect to the field in order to obtain the polarization and the susceptibility. This quantity is obtained from the fact that the number of electron in each unit cell, $q$ in equation (\ref{qen}), should not depend on the electric field. Therefore $\partial_\el q = 0$ so that
\be
  \mu^{(1)}= \l.\frac{\sum_{\alpha} \l(\partial_\el \en^\alpha\r) n_F'(\en^\alpha)}{\sum_{\alpha} n_F'(\en^\alpha)}\r\vert_{\el=0}
\ee
Using this result in the above expression for the susceptibility, we recover equation (\ref{susc}). More generally, $\partial_\el q=0$ gives $\partial_\el \tilde{\mu}(\el) = \frac{\sum_{\alpha} \l(\partial_\el \en^\alpha(\el)\r) n_F'(\en^\alpha(\el)-\tilde{\mu}(\el))}{\sum_{\alpha} n_F'(\en^\alpha(\el)-\tilde{\mu}(\el))}$.

\section{Electron interactions and global versus local filling}
\label{appendix2}
In this appendix, we justify the assumption of a charge distribution that retains the Bravais lattice periodicity even in the presence of an electric field. We therefore relax the local chemical potential hypothesis (which states that the chemical potential depends on the unit cell $n$ through $\mu_n = \tilde\mu(\el) + \el n$) and turn back to a unique global chemical potential $\mu$. The latter serves to impose overall charge neutrality (global) but not necessarily electro-neutrality in each unit cell (local). The new ingredient is to add interactions between charges (electrons and ions) giving a cost to charge inhomogeneities. The goal is to show that a local electro-neutrality within each unit cell naturally emerges in the limit of strong electrostatic interactions.

The WSL spectrum is not bounded from below, and as so, when we write the partition function, we sum over configurations of infinitely negative energy. Such configurations correspond to charge distributions that are highly inhomogeneous: most of the electrons are on one side of the crystal. However, such an electronic filling should have a cost. What would be the influence on the polarization and susceptibility of such an electrostatic cost? Below, we propose a toy model of interacting electrons. It is exactly solvable as it can be seen as a model of independent unit cells.

We consider spinless electrons in a one dimensional two-band system, whose two WSL are $\en^+(\el) + \el n = \en(\el) + \el n$ and $\en^-(\el) + \el n = -\en(\el) + \el n$ (our toy model is assumed to posses a particle-hole symmetry). As WS states are localized, we may associate each state to a unit cell through its center. Every unit cell can then have four states: either empty, or occupied by one electron in either the $\en^-$ or the $\en^+$ ladder, or doubly occupied. For the sake of simplicity, we set an extra cost $2U$ to a doubly occupied unit cell. Hence, the 4 possible ``grand canonical energy levels'', including the ionic contribution, are
\ba
  -\el n, \ \en^- - \mu, \ \en^+-\mu, \nonumber \\
  \en^- + \en^+ +\el n + 2U - 2\mu
\ea
The grand canonical partition function is therefore
\be
  \Xi=\prod_n (e^{\beta \el n}+e^{\beta(\en(\el)+\mu)}+e^{-\beta(\en(\el)-\mu)}+e^{-\beta(\el n + 2U -2\mu)})
\ee
The unit cells range from $n=-(N-1)/2$ to $+(N-1)/2$ with $N$ odd, the system contains $N$ electrons and from $\beta^{-1}\partial_\mu \ln \Xi = N$, we find that the chemical potential is $\mu = U$ at all orders in the field (due to particle-hole symmetry, which is best seen if the electrostatic cost of $2U$ is shared equally by the empty and doubly occupied states). We therefore obtain a simple expression for the free energy $\free = \beta^{-1}\ln \Xi + \mu N$:
\be
  \free = -\frac{N\ln2}{\beta}  -\frac{1}{\beta} \sum_n \ln\left[e^{-\beta U}\cosh(\beta \el n) + \cosh\l(\beta\en(\el)\r)\right]
\ee

From this we are able to compute the polarization as
\be
  \pol = \partial_\el \en \frac{\sinh(\beta\en)}{e^{-\beta U} + \cosh(\beta\en)}
\ee
If the limit of weak interactions $U\ll T$, we can rewrite the polarization as
\be
  \pol = \partial_\el \en \frac{\sinh(\beta\en)}{1 + \cosh(\beta\en)}=-\sum_{\alpha=\pm}(\partial_\el \en^\alpha) n_F\l(\overline{\en^\alpha}\r)
\ee
which agrees with equation (\ref{pol}). Whereas in the limit of strong interactions $U\gg T$, we find
\be
 \pol = \partial_\el \en \tanh(\beta\en)
\ee

At first sight, this is puzzling. We have devised a model in order to show that the local chemical potential hypothesis is valid in the strong interaction limit and we recover our previous results -- obtained using local chemical potentials -- in the opposite limit of weak interactions! The reason is twofold. First, the local chemical potential hypothesis does not play a role at first order in the field (i.e. for the polarization) and only appears at the second order (i.e. for the susceptibility, see below). This is the reason why the polarization at $U=0$ is the same whether one uses a global chemical potential $\mu$ or local chemical potentials $\{\mu_n\}$. Second, in the strong interaction limit, our model freezes so strongly the charge fluctuations that it is actually equivalent to a local {\it canonical ensemble} (exactly one electron in each unit cell) rather than a local {\it grand canonical ensemble} (one electron on average per unit cell). Indeed, the partition function for a single unit cell at $n=0$ occupied by one electron is $Z_1=2\cosh (\beta E(\el))$ giving a polarization $T\partial_\el \ln Z_1 = \partial_\el \en \tanh(\beta\en)$. In the main part of the paper, we developed a local grand canonical approach. However, in the remaining of this appendix, we continue the investigation of the interacting model that resembles a local canonical (rather than grand canonical) ensemble in the strong interaction limit. The aim is to see whether a strong interaction limit is equivalent to imposing a local electronic filling (either in a local canonical or a local grand canonical ensemble).

Taking another derivative, we access the susceptibility
\ba
\chi & = & -\partial^2_{\el} \en \frac{\sinh (\beta E)}{e^{-\beta U}+\cosh(\beta E)}\nonumber \\
&-& (\partial_\el \en )^2 \beta \frac{1+e^{-\beta U}\cosh (\beta E)}{(e^{-\beta U}+\cosh(\beta E))^2}\nonumber \\
&-&\beta \frac{e^{-\beta U}}{e^{-\beta U}+\cosh(\beta E)}\frac{N^2}{12}
\ea
where we used that $\sum_{n=-(N-1)/2}^{(N-1)/2}n^2/N\approx N^2/12$ when $N\gg 1$. On the one hand, in the weak interaction limit $\beta U \ll 1$, we find
\ba
\chi  &=&  -\sum_\alpha \left[\partial^2_{\el} \en^\alpha n_F(\overline{\en^\alpha})+(\partial_\el \en^\alpha)^2 n_F'(\overline{\en^\alpha})\right. \nonumber \\
&+&\left. \frac{N^2}{12}n_F'(\overline{\en^\alpha})\right]
\ea
The two first terms are expected (compare with equation (\ref{suscfinal}) when $\mu^{(1)}=0$, as here $\tilde{\mu}(\el)=\mu=U$ is field independent) but not the last term (proportional to $N^2$). It is not intensive and diverges in the thermodynamic limit. It reflects the fact that imposing a global electronic filling in the presence of an electric field and in the absence of a repulsion between electrons, the system does not remain a band insulator but contains partially filled bands due to inter-band tunneling. Such a conducting systems does not have a finite electric susceptibility in the thermodynamic limit. This is a signature of a metallic behavior (usually best captured at finite frequency). On the other hand, in the strong interaction limit $\beta U \gg 1$, we find
\ba
\chi  &=&  \partial^2_{\el} \en \tanh (\beta E)+(\partial_\el \en)^2 \beta \textrm{sech}^2(\beta E) \nonumber \\
&+& \frac{N^2}{12}e^{-\beta U}\beta \textrm{sech}(\beta E)
\ea
We now see that the term which depends on the size of the crystal is proportional to $N^2 e^{-\beta U}$ and that is is linked to the configurations which present charge inhomogeneity. It can still be controlled in a more stringent limit of strong interactions involving the size of the system. The temperature should only be lower than $U/\ln N$ in order for the last term to be negligible. Typically $U\sim \frac{e^2}{4\pi \epsilon a}$ is of the order of $10$ eV i.e. of $10^5$ K. Therefore even for $N\sim 10^{23}$, the temperature should be lower than $U/\ln N \sim 10^3$ K. When the last term is negligible, the result for the susceptibility is the same as the one that would be obtained from a local canonical ensemble in the absence of interactions. Indeed $T\partial_\el^2 \ln Z_1 |_{\el=0}=\partial^2_{\el} \en \tanh (\beta E)+(\partial_\el \en)^2 \beta \textrm{sech}^2(\beta E)$.

To summarize, we find that, for both the polarization and the susceptibility, one may consider non-interacting electrons provided the charge neutrality is imposed locally in each unit cell rather than globally over the whole crystal. We also see that there is a slight difference between imposing this local electronic filling per unit cell in the canonical or in the grand canonical ensemble (the difference in 1 dimension is due to the small number of electrons involved). In the main part of the article, we assumed a local chemical potential and therefore used a local grand canonical ensemble. \\


This simple toy-model can be extended to the two-dimensional case. Again, we assume a band insulator coming from a two-bands tight-binding model on a lattice. We also assume that the electric field lies along one of the Bravais vector (we call this direction the parallel direction). In this particular case, the crystal retains its translational invariance in the perpendicular direction without having to change the unit cell, so we still have two bands in the perpendicular direction. As we assumed two bands in the parallel direction without electric field, we now have two WSL whose energies are $\en^{\pm}(k_\perp,\el) + \el n_\parallel$. Given a unit cell $n_\parallel$, all the WS states located in that unit cell are Bloch plane waves in the perpendicular direction and confined in the parallel direction, so we set the interaction cost to $U (N_{n_\parallel}-N_0)^2$, where $N_{n_\parallel}$ is the number of electrons on the rung and $N_0$ is the number of electrons needed to realize charge neutrality.

In the analytically-tractable case of flat bands, and in the $N_0\to\infty$ (thermodynamic limit in the perpendicular direction), we note the following facts: (i) when the interaction is set to $U=0$, as soon as the chemical potential reaches the upper band energy, the net charge of each rung diverges as expected; (ii) as soon as we consider $U>0$, whatever the value of the chemical potential, the net charge of the rung remains finite even in the thermodynamic limit, that is, each rung remains very close to charge neutrality.

Hence, this (overly-simplified) two-dimensional toy-model tells us that interactions are likely to enforce local neutrality in the crystal, and supports our approach of neglecting Zener tunneling and enforcing local neutrality at the scale of the unit cell. 


\section{Finite temperature polarization and susceptibility from charge density}
\label{appendix3}
In this appendix, we propose an alternative derivation of the finite temperature polarization and susceptibility starting from the charge density. We consider a \textit{finite} crystalline chain in the presence of an electric field. The energy spectrum $\{\en_\gamma(\el)\}$ (with $\gamma$ representing quantum numbers) is bounded and the eigenstates $\{|\psi_\gamma(\el)\rangle\}$ are well localized (in particular their average position is well defined). In such a case, the polarization can be computed from the charge density in the familiar Clausius-Mossoti approach
\be
\pol = -\frac{1}{N}\int dx x \rho(x)
\ee
(with $a=1$, $e=1$)
where $\rho(x)$ is the total electric charge density (we get rid of the ionic contribution by taking the average ion position as the spatial origin). At finite temperature -- and in a grand canonical picture with global filling fixed by the chemical potential $\mu$ -- it is given by
\be
\rho(x)=\sum_{\gamma} n_F(\en_\gamma(\el)-\mu(\el))|\psi_\gamma(\el,x)|^2
\ee
with $N_{e-}=\sum_{\gamma} n_F(\en_\gamma(\el)-\mu(\el))$. So that the finite field polarization is
\be
\pol(\el)=-\frac{1}{N}\sum_{\gamma} n_F(\en_\gamma(\el)-\mu(\el)) \langle \psi_\gamma(\el)|X|\psi_\gamma(\el)\rangle
\ee
According to the Hellmann-Feynman theorem $\langle \psi_\gamma(\el)|X|\psi_\gamma(\el)\rangle=\langle \psi_\gamma(\el)|\partial_\el \ham|\psi_\gamma(\el)\rangle = \partial_\el \en_\gamma (\el)$ and the finite field polarization becomes:
\be
\pol(\el)=-\frac{1}{N}\sum_{\gamma} n_F(\en_\gamma(\el)-\mu(\el)) \partial_\el \en_\gamma (\el)
\label{ffpol}
\ee
For a finite system, the spontaneous polarization is therefore
\be
\pol=-\frac{1}{N}\sum_{\gamma} n_F(\en_\gamma-\mu) \partial_\el \en_\gamma
\label{spolapp}
\ee
and the susceptibility is
\ba
\chi&=&-\frac{1}{N}\sum_{\gamma} \left[n_F'(\en_\gamma-\mu) (\partial_\el \en_\gamma - \partial_\el \mu)\partial_\el \en_\gamma \right.\nonumber \\
&+&
\left. n_F(\en_\gamma-\mu)\partial_\el^2 \en_\gamma \right]
\label{suscapp}
\ea
with $\partial_\el \mu = [\sum_{\gamma}  n_F'(\en_\gamma-\mu)]^{-1} \sum_{\gamma}  n_F'(\en_\gamma-\mu)\partial_\el \en_\gamma$. We use the convention that $\en_\gamma \equiv \en_\gamma(\el=0)$ and similarly for $\partial_\el \en_\gamma$, $\partial_\el^2 \en_\gamma$ and $\mu$. Note also that $\partial_\el \en_\gamma=\langle \psi_\gamma|X|\psi_\gamma\rangle$ and $\partial_\el^2 \en_\gamma=\langle \partial_\el \psi_\gamma|X|\psi_\gamma\rangle + \langle \psi_\gamma|X|\partial_\el \psi_\gamma\rangle$. If non-degenerate perturbation theory is applicable (which is certainly not the case in the thermodynamic limit and at finite field as there are level crossings), we can further show that $\partial_\el^2 \en_\gamma= 2 \sum_{\delta\neq \gamma} \frac{|\langle \psi_\gamma|X|\psi_\delta \rangle|^2}{E_\gamma - E_\delta}$.

We now would like to take the limit of an infinite chain using our knowledge of the WSL spectrum, equation (\ref{wslp}). An important point is that the zero field limit should be taken after the thermodynamic limit and that there is a discontinuity of the spectrum at zero field. This is due to level crossing when $\el N$ becomes larger than the band gap, which always occur in the thermodynamic limit $N\to \infty$ at any finite field. Therefore, we can not use (\ref{spolapp}) and (\ref{suscapp}) but rather go back to the finite field polarization (\ref{ffpol}) and replace $\gamma \to (\alpha,n)$, $\en_\gamma(\el) \to \en^\alpha_n(\el)=n\el+\en^\alpha(\el)$ so that $\partial_\el \en_\gamma \to \partial_\el \en^\alpha_n = n+\partial_\el \en^\alpha$ and $\partial^2_\el \en_\gamma \to \partial^2_\el \en^\alpha_n = \partial^2_\el \en^\alpha$, leading to
\be
\pol=-\sum_{\alpha} n_F(\overline{\en^\alpha}-\mu)\partial_\el \en^\alpha
\label{spolapp2}
\ee
and
\ba
\chi&=&-\sum_{\alpha} \left[n_F'(\overline{\en^\alpha}-\mu) (\partial_\el \en^\alpha - \partial_\el \mu)\partial_\el \en^\alpha \right.\nonumber \\
&+&
\left. n_F(\overline{\en^\alpha}-\mu)\partial_\el^2 \en^\alpha \right] -\frac{N^2}{12}\sum_{\alpha} n_F'(\overline{\en^\alpha}-\mu)
\label{suscapp2}
\ea
where we used that $N^{-1}\sum_{n=-(N-1)/2}^{(N-1)/2}=1$, $N^{-1}\sum_{n}n=0$ and $N^{-1}\sum_{n}n^2\approx N^2/12$, and where $\partial_\el \mu = [\sum_{\alpha}  n_F'(\overline{\en^\alpha}-\mu)]^{-1} \sum_{\alpha}  n_F'(\overline{\en^\alpha}-\mu)\partial_\el \en^\alpha$. The last term in (\ref{suscapp2}) (proportional to $N^2$) is present because we only imposed a global electronic filling and not a local one. It is a signal that if one waits long enough, a band insulator in a finite field does not remain insulating but becomes a conductor due to inter-band tunneling. This term should therefore be ignored when computing the susceptibility of an insulator. See the corresponding discussion in Appendix \ref{appendix2}, which shows how this term is killed by electrostatic interactions.


If the above replacements $\gamma \to (\alpha,n)$, etc. are made in (\ref{spolapp}) and (\ref{suscapp}), the first derivative becomes
\be
  \partial_\el \en_\gamma = \to \partial_\el \en_n^\alpha = \psop{w_n^\alpha}{X}{w_n^\alpha} = n + \frac{Z^\alpha}{2\pi}
\ee
which is the correct value, while for the second order
\ba
       \frac{1}{2}\partial^2_\el \en_\gamma & \to & \sum_{\beta \neq \alpha} \frac{\abs{\psop{w_n^\alpha}{X}{w_n^\beta}}^2}{\overline{\en^\alpha}-\overline{\en^\beta}} \nonumber \\
  \neq \frac{1}{2}\partial^2_\el \en_n^\alpha & = & \sum_{\beta \neq \alpha}\int_{BZ}\frac{\bco^{\alpha\beta}(k)\bco^{\beta \alpha}(k)}{\en^\alpha(k)-\en^\beta(k)} \\
\ea
The first derivative is correct but not the second. This illustrates the failure of non-degenerate perturbation theory in the case of an infinite crystal. Also we see, that if we use the correct expression for the WSL given in equation (\ref{wslp}), it is possible to compute the polarization and susceptibility in the thermodynamic limit starting from the charge density.


\begin{thebibliography}{99}
\bibitem{Ashcroft}N.W. Ashcroft and N.D. Mermin, \emph{Solid State Physics}, Saunders, Philadelphia (1976), chapter 27.

\bibitem{KSV}R.D. King-Smith and D. Vanderbilt, Phys. Rev. B \textbf{47}, 1651(R) (1993); Phys. Rev. B \textbf{48}, 4442 (1993).

\bibitem{Resta1994}R. Resta, Rev. Mod. Phys. \textbf{66}, 899 (1994).

\bibitem{VanderbiltResta} R. Resta and D. Vanderbilt, Physics of Ferroelectrics: a Modern Perspective, Topics in Applied Physics 105 (Springer ed.), pages 21-68 (2007).

\bibitem{Spaldin} N.A. Spaldin, J. Solid State Chem. \textbf{195}, 2 (2012).

\bibitem{Zak1989}J. Zak, Phys. Rev. Lett. \textbf{62}, 2747 (1989).

\bibitem{Raoux2015}A. Raoux et al., Phys. Rev. B \textbf{91}, 085120 (2015).

\bibitem{Wannier1960}G.H. Wannier, Phys. Rev. \textbf{117}, 432 (1960).

\bibitem{Zak1968}J. Zak. Phys. Rev. Lett. \textbf{20}, 1477 (1968).

\bibitem{Resta2000}R. Resta, J. Phys.: Condens. Matter \textbf{12}, R107 (2000).

\bibitem{Niu2010}Di Xiao, Ming-Che Chang and Qian Niu, Rev. Mod. Phys. \textbf{82}, 1959 (2010).

\bibitem{NunesVanderbilt}R.W. Nunes and D. Vanderbilt, Phys. Rev. Lett. \textbf{73}, 712 (1994).

\bibitem{Gonze}R.W. Nunes and X. Gonze, Phys. Rev. B \textbf{63}, 155107 (2001).

\bibitem{Souza}I. Souza, J. Iniguez and D. Vanderbilt, Phys. Rev. Lett. \textbf{89}, 117602 (2002).

\bibitem{Umari}P. Umari and A. Pasquarello, Phys. Rev. Lett. \textbf{89}, 157602 (2002).

\bibitem{Kirtman}M. Springborg and B. Kirtman, Phys. Rev. B \textbf{77}, 045102 (2008).

\bibitem{SpringborgKirtman}B. Kirtman, M. Ferrero, M. R\'erat and M. Springborg, J. Chem. Phys. \textbf{131}, 044109 (2009).

\bibitem{Nourafkan2013}R. Nourafkan and G. Kotliar, Phys. Rev. B \textbf{88}, 155121 (2013).

\bibitem{SwiecickiSipe} S.D. Swiecicki and J.E. Sipe, Phys. Rev. B \textbf{90}, 125115 (2014).

\bibitem{footnote1}We make the hypothesis that the electric field seen by ions and electrons is identical to the macroscopic field present in the crystal. In other words, we neglect the difference between the local field and the macroscopic field.

\bibitem{footnote2}The choice of this gauge is motivated by the fact that we want to access the energies of the system through the Hamiltonian. In the time-dependent vector potential gauge $\boldsymbol{A}=-t \boldsymbol{\el}$, any static electric field would lead to a time dependence through the Peierls substitution, while in the scalar gauge $A_0=-\el X$ the Hamiltonian is time independent and therefore corresponds to the energy.

\bibitem{footnote4}It is therefore important to understand that $\alpha$ is a band index only at $\el =0$. But as soon as $\el \neq 0$, $\alpha$ becomes a WSL or ladder index.

\bibitem{Leo}K. Leo, Semicond. Sci. Technol. \textbf{13}, 249 (1998).


\bibitem{footnote3}Bloch eigenvectors are defined up to a gauge choice, which is the phase they come with, and two different choice of phases will lead to two different set of Wannier functions. By suitable choice of phases, we mean choices of phases such that the Berry connection is continuous.

\bibitem{LandauLifshitz8}L.D. Landau and E.M. Lifshitz, \emph{Electrodynamics of continuous media}, Volume 8 of Course of Theoretical Physics, (Pergamon Press, 1960), \S 14.

\bibitem{RiceMele1982}M.J. Rice and E.J. Mele, Phys. Rev. Lett. \textbf{49}, 1455 (1982).

\bibitem{SSH}W. P. Su, J. R. Schrieffer and A. J. Heeger, Phys. Rev. Lett. \textbf{42}, 1698 (1979).


\bibitem{ChenLee2011a} K.-T. Chen and P.A. Lee, Phys. Rev. B \textbf{84}, 113111 (2011).

\bibitem{footnote5}In 1D, the maximally localized Wannier function is also obtained as the eigenstate of the projected position operator onto the occupied band, see \cite{Kivelson1982}.

\bibitem{Kivelson1982}S. Kivelson, Phys. Rev. B \textbf{26}, 4269  (1982).

\bibitem{MarzariRMP2012}Nicola Marzari, Arash A. Mostofi, Jonathan R. Yates, Ivo Souza, and David Vanderbilt, Rev. Mod. Phys. \textbf{84}, 1419 (2012)

\bibitem{ChenLee2011b} K.-T. Chen and P.A. Lee, Phys. Rev. B \textbf{84}, 205137 (2011).

\bibitem{fredpunpublished}F. Pi\'echon, unpublished.

\end{thebibliography}

\end{document}